\documentclass[aps,prd,twocolumn,superscriptaddress,showpacs,amsmath,amssymb,nofootinbib]{revtex4-2}

\usepackage{graphicx} 
\usepackage{epstopdf} 
\usepackage{dcolumn} 
\usepackage{xcolor} 
\usepackage{amssymb} 
\usepackage{hyperref} 
\usepackage[utf8]{inputenc} 
\usepackage[english]{babel} 
\usepackage[displaymath, mathlines]{lineno} 
\usepackage{blindtext} 
\usepackage{comment}
\usepackage{tikz-feynman}
\usepackage{amsmath,bm}
\usepackage{multirow}
\usepackage{booktabs}
\usepackage{subfigure,subfigmat}
\usepackage{orcidlink}  

\graphicspath{{figures/}} 

\usepackage{enumitem}  
\usepackage{feynmf}    

\input belle2sym.tex

\begin{document}


\title{Unlocking time-dependent \textbf{\textit{CP}} violation without signal vertexing at \textbf{\textit{B}} factories}

\author{M. Dorigo\,\orcidlink{0000-0002-0681-6946}}
\affiliation{INFN Sezione di Trieste, Trieste, Italy}

\author{S. Raiz\,\orcidlink{0000-0001-7010-8066}}
\affiliation{Deutsches Elektronen-Synchrotron (DESY), Hamburg, Germany}

\author{D. Tonelli\,\orcidlink{0000-0002-1494-7882}}
\affiliation{INFN Sezione di Trieste, Trieste, Italy}

\author{R. \v{Z}leb\v{c}\'{i}k\,\orcidlink{0000-0003-1644-8523}}
\affiliation{INFN Sezione di Trieste, Trieste, Italy}

\begin{abstract}
We present a method to measure time-dependent \cp violation in \Bz decays produced in $\FourS \to \Bz\Bzb$ events at $B$ factories  without reconstructing the signal decay vertex. 
The method exploits the sensitivity of the tag \Bzb  decay time to \cp violation in the \Bz signal. It relies on a compact \epem interaction region and excellent vertex resolution, which enable precise measurement of the displacement between production and decay vertices of the tag meson.  
We study an application to $\Bz\to\piz\piz$ decays using a simplified simulation  that approximates the conditions of the Belle II experiment at the SuperKEKB collider. Our approach 
achieves a sensitivity on mixing-induced \cp violation that would require 20~times larger samples with standard approaches. When incorporated into the $B \to \pi\pi$ isospin analysis, the expected results would reduce the degeneracy of $\phi_2$ solutions and significantly increase precision.  This work opens a path to previously inaccessible \cp-violation studies enhancing and accelerating the reach of Belle II and future flavor physics programs.

\end{abstract}

\maketitle

\section{Introduction}
Measurements of decay-rate asymmetries that violate the combined transformation of charge conjugation and parity (\cp symmetry) are central to the study of the weak interactions of quarks. These measurements provide stringent tests of the Cabibbo-Kobayashi-Maskawa (CKM) mechanism, which prescribes that all \cp-violating phenomena in the standard model of particle physics are accommodated by a single complex phase in the quark couplings with $W^\pm$ bosons~\cite{Kobayashi:1973fv}. For neutral strange, charm, and bottom mesons, \cp violation phenomenology is further enriched by particle-antiparticle oscillations, yielding \cp-violating decay-rate asymmetries that depend on meson decay time. The observation of these asymmetries in the $\Bz \to \jpsi \Kz$ decays provided the conclusive experimental confirmation of the CKM paradigm in the early 2000s~\cite{BaBar:2001pki,Belle:2001zzw}. The focus then shifted toward using time-dependent \cp violation as a sensitive probe for physics beyond the standard model (see, {\it e.g.},  Ref.~\cite{Fleischer:2024} and references therein).

Bottom mesons have been extensively studied at experiments in energy-asymmetric electron-positron collisions near the $\Upsilon(4S)$ resonance (so-called $B$~factories), such as BaBar, Belle, and \belletwo,  as well as hadron-collider experiments, such as CDF, D0,  LHCb, CMS, and ATLAS~\cite{Banerjee:2024znd}. Two ingredients are essential for measuring time-dependent \cp asymmetries: determining whether the oscillating meson is a $B^0$ or a $\overline{B}^0$ ({\it i.e.}, flavor) at a known time before decay, and measuring the proper-time interval between that instant and the decay. The experimental approaches to flavor determination---and thus to establishing the relevant time---differ significantly between $B$~factories and hadron colliders.

At hadron-collider experiments, $B^0$ and $\overline{B}^0$ mesons originate from the hadronization of incoherently produced $b\bar b$ quark pairs. Flavor tagging is performed at production (see, {\it e.g.}, Refs~\cite{CDF:2000vye,Aaij:2012mu, Aaij:2017lff}) and the \cp-violating decay-rate asymmetry depends only on the signal decay time, which is measured from the observed distance between its production and decay positions and  its momentum. 

At $B$~factories, neutral bottom mesons are produced in quantum-entangled $\Bz\Bzb$ pairs through the process $\epem\to\FourS\to \Bz\Bzb$ , where the $\ep$ and $e^-$ beams have different energies, boosting the $\FourS$ in the laboratory frame.  The entangled pair evolves coherently until one meson decays, at which time the two flavors are opposite. Then, the remaining meson evolves  until it decays. If one meson, identified as $\Btag$, decays into a final state that unambiguously defines its flavor and the other, $\BCP$, decays into a final state accessible to both flavors, a measurement of time-dependent \cp violation for $\BCP$ is possible  (see, {\it e.g.}, Ref.~\cite{Bevan:2014iga}). 
The corresponding time-evolution and \cp-violating decay-rate asymmetry are described in terms of the flavor correlations between the two mesons and 
\begin{equation}
\deltat = \tcp - \ttag, 
\end{equation}
the difference between the two decay times, which can be either positive or negative. This difference is measured from the spatial separation of the two decay positions using the known boost. 

In this paper, we present a method to measure time-dependent \cp-violating asymmetries at $B$~factories for decays in which $\deltat$ is not available because the $\BCP$ decay position is not reconstructed. This limitation arises when the final state includes only neutral particles such as photons or $K^0$ mesons as in $\Bz \to \piz\piz$, $\Bz \to K^0 \piz(\piz)$, $\Bz \to K^0 \Kzb(K^0)$, or the yet-unobserved $\Bz \to \gamma\gamma$ decays. Similar ideas have been explored in the past~\cite{Foland:1999wq}. 

In $\Bz$ decays involving neutral pions, the $\piz$ is typically reconstructed via $\piz \to \gamma\gamma$, which provides no vertex information. Processes like $\piz \to \epem \gamma$, whether from Dalitz decays or photon conversions, enable signal-vertex reconstruction, but their low branching fractions, $\mathcal{O}(10^{-2})$, make measurements of time-dependent \cp violation feasible only with extremely large data samples.  In $\Bz$ decays involving neutral kaons, no $\KL$ decays and only the fraction of $\KS \to \pi^+\pi^-$ decays that occur in the inner volume of silicon-vertex detectors provides precise signal-vertex information. In $\Bz$  decays into photons, only conversions can be used, resulting in severe sample-size limitations.

Our approach overcomes these limitations, enabling sensitive measurement of time-dependent \cp violation in samples that are already available. 
 The approach relies solely on determining $\ttag$, from the displacement between the $\Btag$ production and decay points, in analogy with decay-time reconstruction in hadron collisions. However, it exploits the intrinsic connection of \ttag to the \BCP asymmetry due to the quantum-entangled nature of the $\Bz\Bzb$ pair characteristic of $B$~factories.

Key to our method is the resolution of the measurement of the \Btag production and decay positions. The latter depends on detector design and performance and it is important because it contributes to the precision on the measurement. However, the former, which depends on the features of the \epem interaction region (IR), is a dominant limiting factor. The entire feasibility of the method is driven by the following collider-specific enablers:  
a small IR transverse size, a nonzero crossing angle for the beams, and a sufficient \FourS boost.

We first describe the approach with an analytical derivation based on the standard formalism of time-evolution of quantum-entangled $\Bz\Bzb$ pairs. We then use simplified simulated experiments that approximate the experimental conditions of the \belletwo\ experiment at the SuperKEKB collider to demonstrate quantitatively the performance with a specific case of interest: the measurement of the time-dependent \cp asymmetry in $\Bz\to\piz\piz$ decays. 
Natural units are used, with $\hbar =c=1$, throughout the paper. 

\begin{figure*}[tb]
\includegraphics[width=0.4\linewidth] {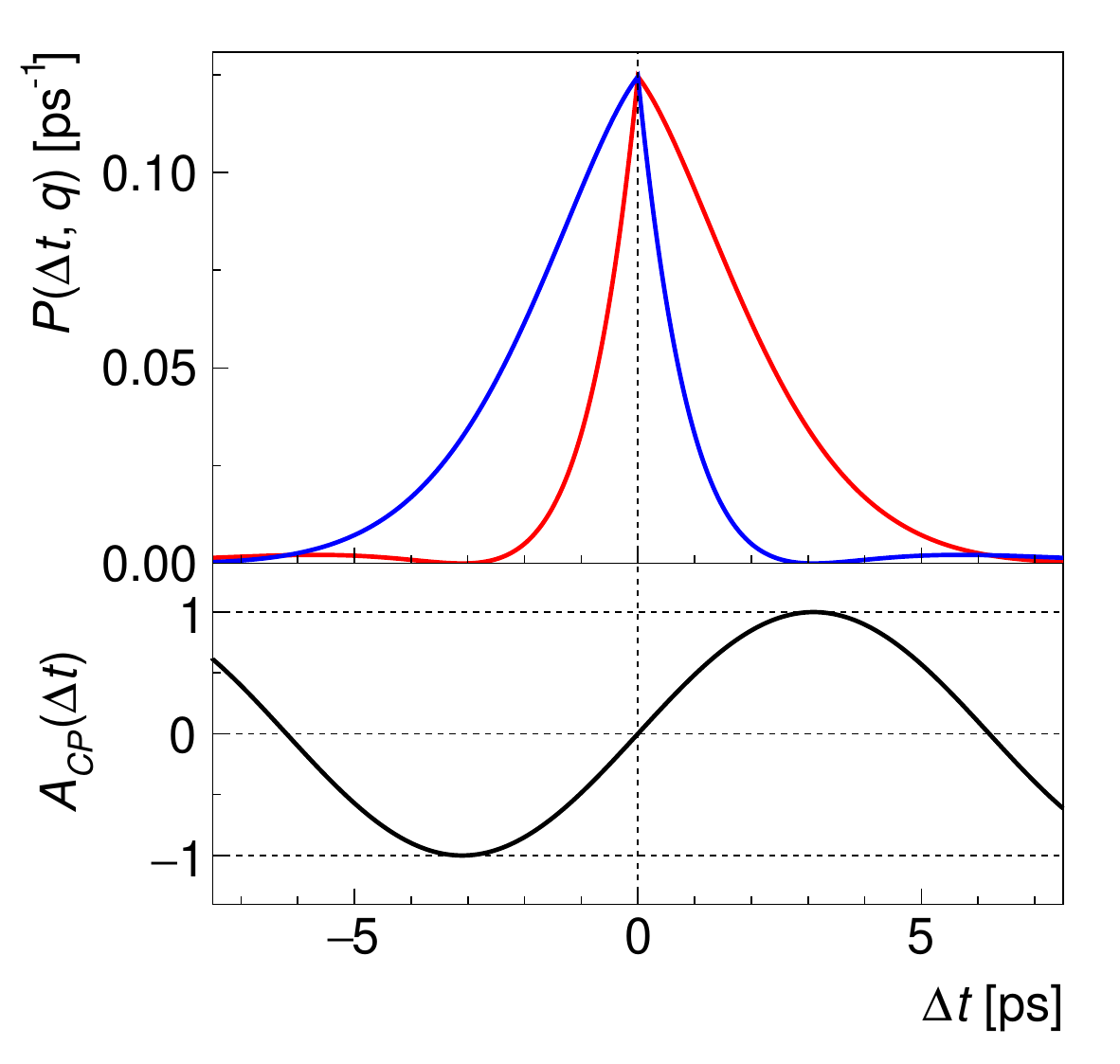} \hspace{1.5cm}
\includegraphics[width=0.4\linewidth]{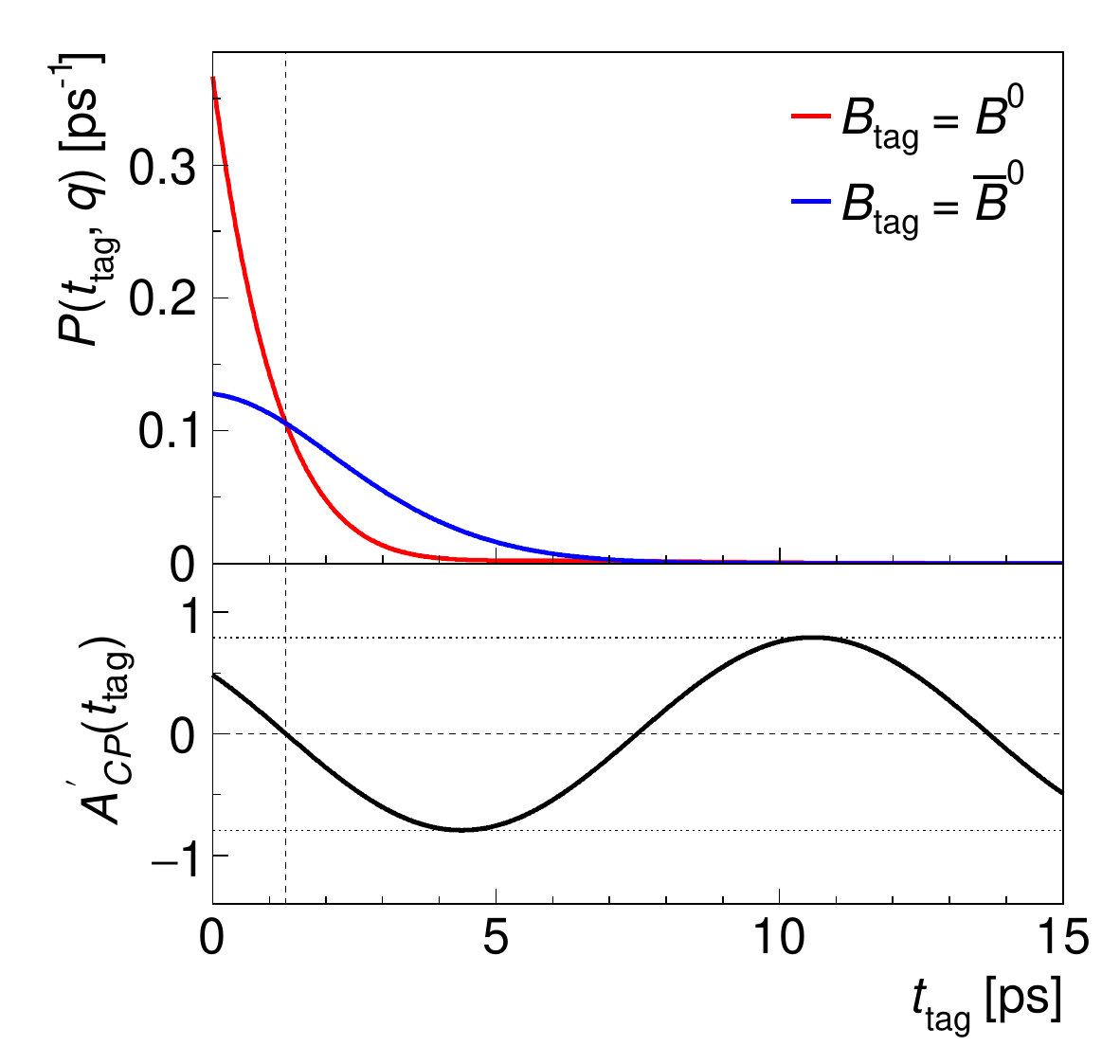}
\caption{(Top panels) Theoretical decay rates as a function of (left) \deltat and (right) \ttag, for tagged (blue) $\Bz$ and (red) $\Bzb$ mesons assuming $S=1.0$ and $C=0.0$. (Bottom panels) corresponding \cp-violating decay-rate asymmetries (left) $\mathcal{A}_{\cp}(\deltat)$ and (right) $\mathcal{A}_{\cp}'(\ttag)$. The horizontal dashed lines show the maximum oscillation amplitudes.  The vertical dashed line shows (left) the \deltat origin and (right) the $\hat{t}$ value in Eq.~(\ref{eqn_t0}). The decay time \ttag is non-negative by definition.}
\label{fig:truthTimes}
\end{figure*}

\section{Time-dependent asymmetry from the tag decay time}
In a $B$~factory experiment, the time-dependent decay rate into a \cp eigenstate of a neutral $B$ meson  is~\cite{Bevan:2014iga}
\begin{linenomath}
\begin{align}\label{eqn_dt1}
\begin{split}
\mathcal{P}(\Delta t,q) = \frac{e^{-|\Delta t|/\tau}}{4\tau} \biggl(1 + q\bigl[ \scp\sin\Delta m^{} \Delta t \\
- \ccp\cos\Delta m^{}\Delta t  \bigr] \biggr)\,,
\end{split}
\end{align}
\end{linenomath}
where we assume $\cp T$ invariance ($T$ being the time-reversal transformation) and neglect a small lifetime difference between the two $B$ physical states, which have a mass difference $\Delta m$ and lifetime $\tau$. The signal decay rate differs for \Bz and \Bzb according to the \Btag flavor $q$ at $\deltat =0$\,ps. The flavor $q$ equals either $+1$ for $\Btag=\Bz$ or $-1$ for $\Btag=\Bzb$. 
Such a difference generates a nonzero time-dependent \cp-violating asymmetry, 
\begin{linenomath}
\begin{align}\label{eqn_acp}
\begin{split}
\mathcal{A}_\cp(\Delta t) &=  \frac{\mathcal{P}(\Delta t,+1) - \mathcal{P}(\Delta t,-1)}{\mathcal{P}(\Delta t,+1) + \mathcal{P}(\Delta t,-1)}  \\
& =\scp\sin\Delta m^{} \Delta t 
- \ccp\cos\Delta m^{}\Delta t.
\end{split}
\end{align}
\end{linenomath}
when either of the \cp-violating coefficients \scp or \ccp, which are functions of the relevant combinations of quark couplings involved in the transition, differ from zero.

Our method relies solely on the measurement of \ttag. We write the probability of observing a 
signal meson  decaying at time \tcp accompanied by a partner \Btag with flavor $q$ decaying at time \ttag, with both times measured relative to the instant of the \FourS decay~\cite{Foland:1999wq,BaBar:1998yfb}, 
\begin{linenomath}
\begin{align}\label{eqn_t2D}
\begin{split}
\mathcal{P}(\ttag, \tcp, q) =  \frac{e^{-\frac{\tcp+\ttag}{\tau}}}{2\tau^2} 
\biggl(&1 + q\bigl[ \scp\sin \Delta m^{} (\tcp - \ttag) \\
&-\ccp\cos \Delta m^{} (\tcp - \ttag)  \bigr] \biggr)\,,
\end{split}
\end{align}
\end{linenomath}
and then integrate it over the unobserved signal decay-time \tcp to obtain the dependence on \ttag only and express it in a form that resembles Eq.~(\ref{eqn_dt1}),
\begin{linenomath}
\begin{align}\label{eqn_tTag}
\begin{split}
\mathcal{P}(\ttag,q) =  \frac{e^{-\ttag/\tau}}{2\tau} \biggl(1 +
q\bigl[ \scp' \sin \Delta m^{} (\ttag-\hat{t}\,)\\
-\ccp'\cos \Delta m^{} (\ttag-\hat{t}\,) \bigr] \biggr)\,,
\end{split}
\end{align}
\end{linenomath}
in which 
\begin{linenomath}
\begin{align}\label{eqn_t0}
\hat{t} = \frac{1}{\Delta m} \arctan (\Delta m \tau)\approx 1.294\,\rm{ps}.
\end{align}
\end{linenomath}

The corresponding time-dependent \cp asymmetry is
\begin{linenomath}
\begin{align}\label{eqn_acp_ttag}
\begin{split}
\mathcal{A}_\cp'(\ttag) 
 =\scp'\sin \Delta m^{} (\ttag-\hat{t}\,) 
- \ccp'\cos \Delta m^{} (\ttag-\hat{t}\,)\,,
\end{split}
\end{align}
\end{linenomath}
in which
\begin{linenomath}
\begin{align}\label{eqn_SC}
S' = - \frac{S}{\sqrt{1 + (\tau \Delta m)^2}}, \quad C' = \frac{C}{\sqrt{1 + (\tau \Delta m)^2}}\,.  
\end{align}
\end{linenomath}
Figure~\ref{fig:truthTimes} compares the standard decay rate $\mathcal{P}(\Delta t,q)$ and the rate $\mathcal{P}(\ttag,q)$ obtained by only observing  $\ttag$, along with the corresponding asymmetries $\mathcal{A}_\cp(\deltat)$ and $\mathcal{A}_\cp'(\ttag)$, for $S=1.0$ and $C=0.0$. The mass difference and the lifetime are set to the values known for the \Bz meson~\cite{ParticleDataGroup:2024cgo}. The standard \cp asymmetry $\mathcal{A}_\cp(\deltat)$ is reduced to an asymmetry $\mathcal{A}_\cp'(\ttag)$, whose maximum amplitude is smaller by a factor
\begin{equation}
\label{eqn_SC_dumping}
    \frac{1}{\sqrt{1 + (\tau \Delta m)^2}} \approx 0.792\,.
\end{equation}
The advantage of using \deltat, when both \tcp and \ttag are available, is that the amplitude of the time-dependent oscillation is not damped and reaches a maximum earlier in time. However,  time-dependent \cp violation remains accessible, though reduced,  if the decay vertex of \BCP is not reconstructed. The reduction is the combined effect of the dampening of the asymmetry and the shift of its maximum value to higher decay times by $\hat{t}$, which result in a smaller yield.  Note that Eqs.~(\ref{eqn_tTag}) and~(\ref{eqn_acp_ttag}) also apply to \BCP, when replacing \ttag with \tcp and inverting the sign of $S'$.

When measuring time-dependent \cp violation, three experimental effects influence the standard asymmetry $\mathcal{A}_\cp(\Delta t)$ and must also be considered for $\mathcal{A}_\cp'(\ttag)$. Tag-side interference~\cite{Long:2003wq}  is expected to induce a small bias and is therefore neglected in the following. The other two effects reduce the observable asymmetry and are more relevant: the efficiency for assigning the correct flavor and the resolution on the time measurement. 

Considering a fraction $w$ of events with wrong flavor assignment, the amplitude of the time-dependent \cp asymmetry is reduced by a factor $1-2w$. The corresponding dilution of the sensitivity to the asymmetry is expressed by the effective flavor-tagging efficiency, a factor that scales the actual signal-sample size down to the fraction that is effectively sensitive to the decay-rate asymmetry. The effective flavor-tagging efficiency is typically 35\% at $B$~factories~\cite{Bevan:2014iga, Belle-II:2024lwr}. 

\begin{figure}[t]
\includegraphics[width=0.8\columnwidth]{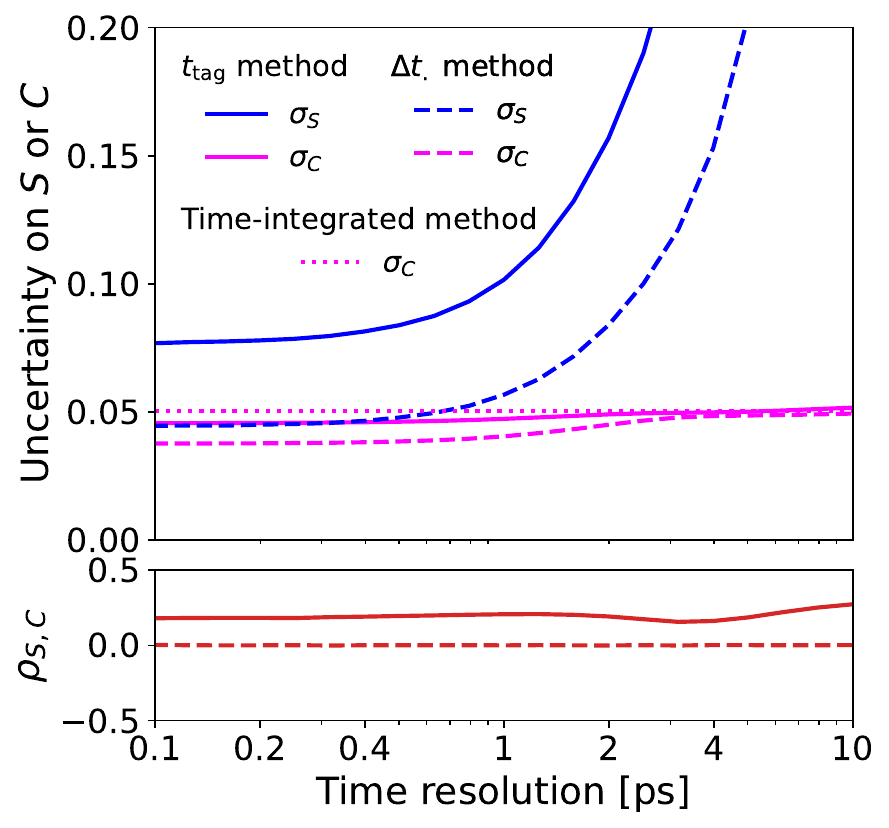}
\caption{(Top panel) uncertainties on the \cp-violating coefficients (blue) $S$ and (magenta) $C$ as functions of the resolution on (solid line) $\ttag$ or (dashed line) $\deltat$, for $10^3$ perfectly-tagged signal-only decays. (Bottom panel) Linear correlation between the coefficients.}
\label{fig:SCresScan}
\end{figure}

 Dilution due to time resolution is the dominant limitation to the sensitivity to $\mathcal{A}_\cp'(\ttag)$ as misreconstructing the time clearly degrades the observed time modulation.  Figure~\ref{fig:SCresScan} shows uncertainties on the \cp-violating coefficients $S$ and $C$ as functions of the resolution on either \ttag, in the case of the asymmetry $\mathcal{A}_\cp'(\ttag)$, or \deltat, in the case of a standard asymmetry $\mathcal{A}_\cp(\deltat)$. To single out the relevant dependence, here we are assuming a signal-only sample of $10^3$ perfectly flavor-tagged decays.  Resolutions are intended as the widths of the relevant Gaussian functions.
All uncertainties increase as the resolution on the relevant time measurement worsens. When both the tag and signal vertices are available, the uncertainties with the $\ttag$-based approach are worse than those with the standard \deltat-based method, because of the dampening effects from Eq.~\ref{eqn_SC_dumping} and the time shift $\hat{t}$ of $\mathcal{A}_{\cp}'(\ttag)$. 
The $C$ uncertainties depend weakly on the time resolution and saturate at the level expected in a time-integrated measurement, which is feasible because, for \Bz mesons, the oscillation period is comparable to their lifetime. As a result, the cosine term in the asymmetry of Eq.~(\ref{eqn_acp}) does not average out. The time-integrated asymmetry is, however, reduced by a factor $1 - 2\chi$, where $\chi = (\Delta m\, \tau)^2/[2(1 + (\Delta m\, \tau)^2)] \approx 0.18$.  
The $S$ uncertainties are nearly constant up to a resolution of about 0.5\,ps. At poorer time resolutions, they deteriorate as expected from the damping of the oscillation amplitude, which is approximately  $\exp[-(\Delta m\, \sigma)^2/2]$, where $\sigma$ is the time resolution~\cite{Moser:1996xf}. Figure~\ref{fig:SCresScan} shows also the linear correlation between $S$ and $C$ as a function of the time resolution. In measurements based on \ttag, the correlation is nonzero (approximately 20\%) because the restriction $0< \ttag < +\infty$ does not allow the cancellation of correlations between negative and positive $\Delta t$ that occurs in the standard method.

The resolution on \deltat is about 0.5\,ps at \belletwo\  and around 1.0\,ps in Belle and Babar~\cite{Belle-II:2024lwr,Tajima:2004kga,Aubert:2002zz}. The resolutions on \ttag are generally expected to be  worse, as they critically depend not only on accurately determining the \Btag decay point, but also its production point, unlike in the standard method.

\begin{figure}[tb]
\includegraphics[width=0.8\columnwidth]{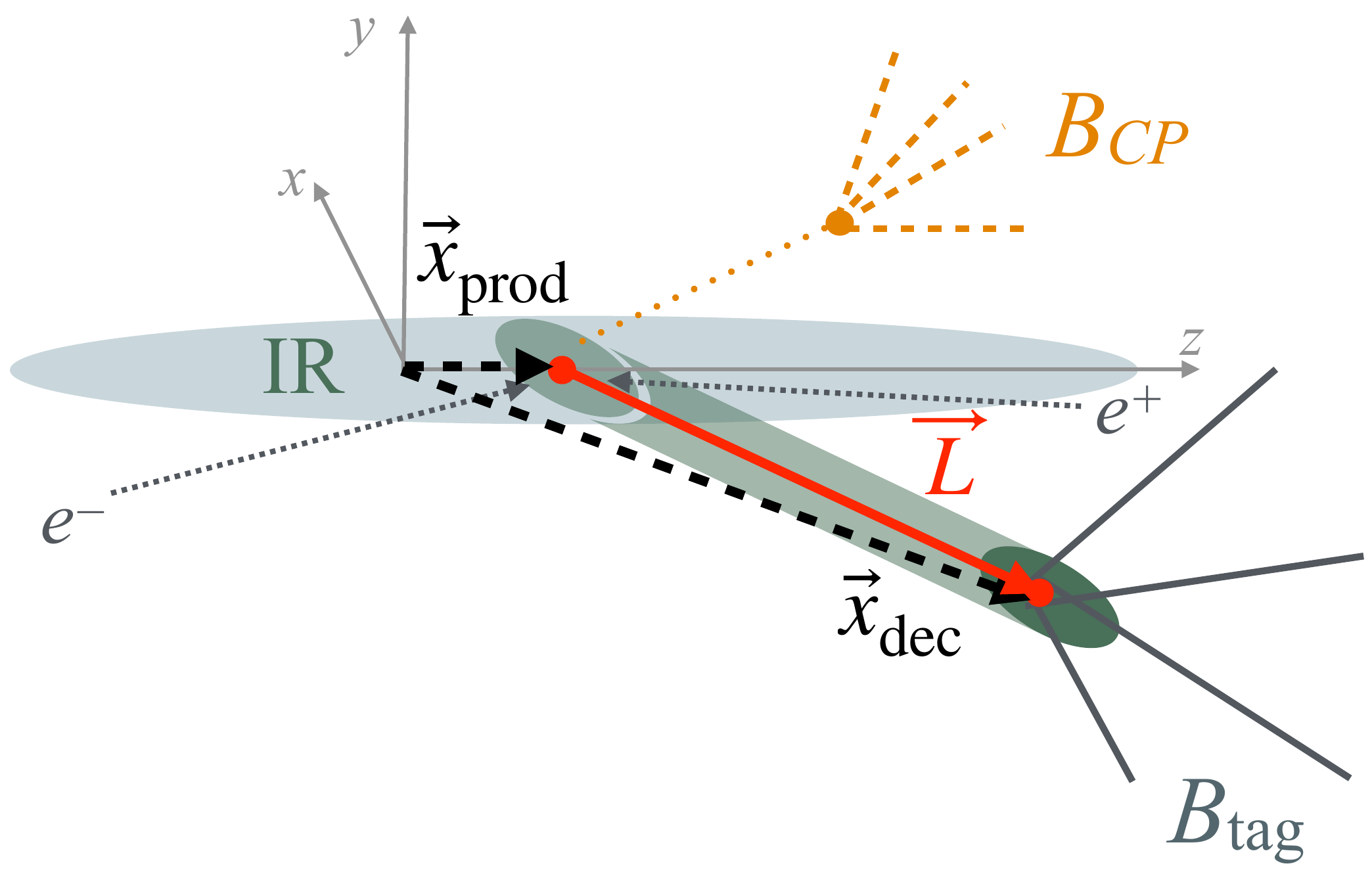}
\caption{Sketch of a $B^0\overline{B}^0$ event topology in the vicinity of the interaction region. Nothing is to scale.} 
\label{fig:BvtxScheme}
\end{figure}

\section{Measurement of the tag decay time}

The production and decay space-points of the \Btag meson identify a flight-distance vector, $\vec{L}$. 
The \Btag decay time 
\begin{equation}
\label{eqn_ttag_def}
    \ttag = m_{B} \frac{\vec{L} \cdot \vec{p}_\textrm{tag}} {|p_\textrm{tag}|^2}\,,
\end{equation}
in which $m_{B}$ is the known \Bz mass, 
is reconstructed by projecting the flight distance onto the $\vec{p}_\textrm{tag}$ momentum direction, to average out minor direction biases due to experimental inaccuracies. In typical measurements of time-dependent \cp violation, the decay of the signal \BCP meson is fully reconstructed, and therefore the momentum $\vec{p}_\cp$ is known. This holds independently of the availability of signal vertex information. The momentum of the partner \Btag meson is therefore inferred as $\vec{p}_\textrm{tag} = \vec{p}_{\FourS} - \vec{p}_\textit{CP}$, where $\vec{p}_{\FourS}$ is the momentum of the \FourS meson in the $\epem\to\FourS\to\Bz\Bzb$ process, which is known precisely from the accelerator parameters.\footnote{If $\vec{p}_\textit{CP}$ cannot be measured, $\vec{p}_\textrm{tag}$ can be approximated by the vector sum of the momenta of the $B_\textrm{tag}$ decay products. However, since $B_\textrm{tag}$ decays are typically reconstructed inclusively to maximize efficiency---and often involve neutrinos---this approximation results in poorer momentum resolution.}  All momenta are measured in the laboratory reference frame. 

The precision on $\vec{L}$ depends on the precision on the \Btag production and decay space-points, and usually drives the precision on the time measurement, as the fractional uncertainty on the momentum is typically much smaller.

The precision on the \Btag production and decay space-points depends on the details of the interaction region. At $B$~factories, the IRs are typically ellipsoidal, with major axes approximately aligned with the beam directions, which define the $z$ axes of the coordinate systems (see Fig.~\ref{fig:BvtxScheme}). The IR spatial dimensions are specific to each collider and are typically determined using processes such as $\epem \to \mu^+\mu^-$~\cite{Kozanecki:2008zz}. They can generally be modeled in each spatial direction by a Gaussian distribution, whose width represents the spatial extent of the IR core along that direction. Typical IR widths for the SuperKEKB (\belletwo), KEKB (Belle), and PEP-II (BaBar) colliders are listed in Table~\ref{tab:BfactoriesPars}. The IR dispersions along the $x$ and $y$ directions are particularly relevant to the precision of $\vec{L}$. 


\begin{table}[t]
\begin{tabular}{lccc}
\hline\hline
                       &    \belletwo & Belle & BaBar \\ 
                     \hline 
$e^-$ ($e^+$) beam energy [GeV]     &  7.0 (4.0)   & 8.0 (3.5)   & 9.0 (3.0)  \\
Beam energy spread [MeV]  &   5.45   &  5.36      &    4.63   \\ 
Crossing angle [mrad]       & 83 &  22   & 0    \\
Interaction region $x$ width [\mum]  &   13 &   70   &  148 \\
Interaction region $y$ width [\mum]   &   0.2 &   1.0  &  6.9  \\
Interaction region $z$ width [\mum] & 350  &   6000   &  15150  \\ 
\Btag vertex $x$-$y$ resolution [\mum]    &   30 &  80   & 80 \\   
\Btag vertex $z$ resolution [\mum]  &   30  &  100    & 125 \\
\hline\hline
\end{tabular}
\caption{Summary of $B$~factory parameters relevant for the measurement of \ttag~\cite{Aubert:2002zz,Ohnishi:2013fma,KEKTsukuba:1995urg,Kozanecki:2008zz,zlebcik_2023_8119739,Belle:2021lzm,BaBar:2004rrm}.}
\label{tab:BfactoriesPars}
\end{table}

The \Btag decay vertex is accurately determined, owing to the collision boost and high-resolution silicon detectors installed at small radial distances from the IR. In the $z$ direction, the \Btag-vertex resolution is typically an order of magnitude, or more, smaller than the IR size (see Table~\ref{tab:BfactoriesPars}). However, the average \Btag flight path---ranging from approximately 130\,\mum at \belletwo\ to 260\,\mum at BaBar---and its near alignment with the $z$ axis, keeps the decay vertex within the IR ellipsoid. Approximating the resolution on the \Btag production point with the IR size would therefore result in a large uncertainty on $\vec{L}$, spoiling the measurement of \ttag.

To determine \ttag, we use instead a topological fit to the coordinates of the measured interaction point $\vec{x}_\mathrm{IR}$ and \Btag decay vertex, $\vec{x}_\textrm{tag}$, using their respective covariance matrices, $V_\mathrm{IR}$ and $V_\textrm{tag}$. The covariance matrix $V_\mathrm{IR}$ is derived from the IR spatial widths, while $V_\textrm{tag}$ is determined from the \Btag vertex resolution.  To mitigate the degradation in precision due to the large IR size, especially along the major axis of the ellipsoid, we incorporate in the fit the known \Btag momentum, $\vec{p}_\textrm{tag}$, which constrains the flight direction and thereby greatly improves the resolution on $\vec{L}$.  
The fit determines the \Btag production and decay positions, $\vec{x}_\mathrm{prod}$ and $\vec{x}_\mathrm{dec}$, by minimizing the $\chi^2$ function
\begin{linenomath}
\begin{align}\label{eqn_chi2}
\begin{split}
\chi^2 &= (\vec{x}_\mathrm{prod} - \vec{x}_\mathrm{IR})^T V_\mathrm{IR}^{-1} (\vec{x}_\mathrm{prod} - \vec{x}_\mathrm{IR})\\
 &+ (\vec{x}_\mathrm{dec} - \vec{x}_\textrm{tag})^T V_\textrm{tag}^{-1} (\vec{x}_\mathrm{dec} - \vec{x}_\textrm{tag})\,,
 \end{split}
\end{align}
\end{linenomath}
in which the decay position 
\begin{equation}
\label{eqn_xdec_fit}
\vec{x}_\mathrm{dec} = \vec{x}_\mathrm{prod} + \frac{t_\mathrm{tag}}{m_B} \vec{p}_\mathrm{tag}    
\end{equation}
is a function of \ttag through the known \Btag momentum. The parameters determined by the fit are $\vec{x}_\textrm{prod}$ and \ttag (see Appendix~\ref{app_formulas} for their analytical derivation).  The uncertainty on $\vec{p}_\textrm{tag}$ is neglected in Eq.~(\ref{eqn_chi2}) because it is irrelevant. However, it can be introduced by adding a constraint on the momentum in the $\chi^2$ function, for example, when the $\vec{p}_\cp$ momentum cannot be measured. 

Geometrically, the \Btag production space-point is identified by the overlap between the IR ellipsoid and a cylinder whose axis is the \Btag momentum and whose radius equals the \Btag-vertex resolution. The center of this overlap marks the \Btag production space-point, and the size of the overlap region defines the resolution (see Fig.~\ref{fig:BvtxScheme}).

\section{Tag decay-time resolution}
To investigate the \ttag resolution expected at $B$~factory experiments, we generate simplified simulated events according to the decay rates of Eq.~(\ref{eqn_tTag}) and using the parameters of the IR, \Btag vertex resolution, and beam energies reported in Table~\ref{tab:BfactoriesPars}, to simulate the relevant conditions of each experiment.  The energies and momentum magnitudes of the $\Bz$ and $\Bzb$ mesons in $\FourS \to \Bz\Bzb$ decays are determined from energy-momentum conservation. Their angular distribution follows the known dependence due to the transverse polarization of the \FourS mesons produced in \epem collisions~\cite{Bevan:2014iga}. We do not simulate geometric detector acceptance and neglect resolution effects on the measurement of the $B$ momentum.

\begin{figure}[tb]
\centering
\includegraphics[width=0.8\columnwidth]{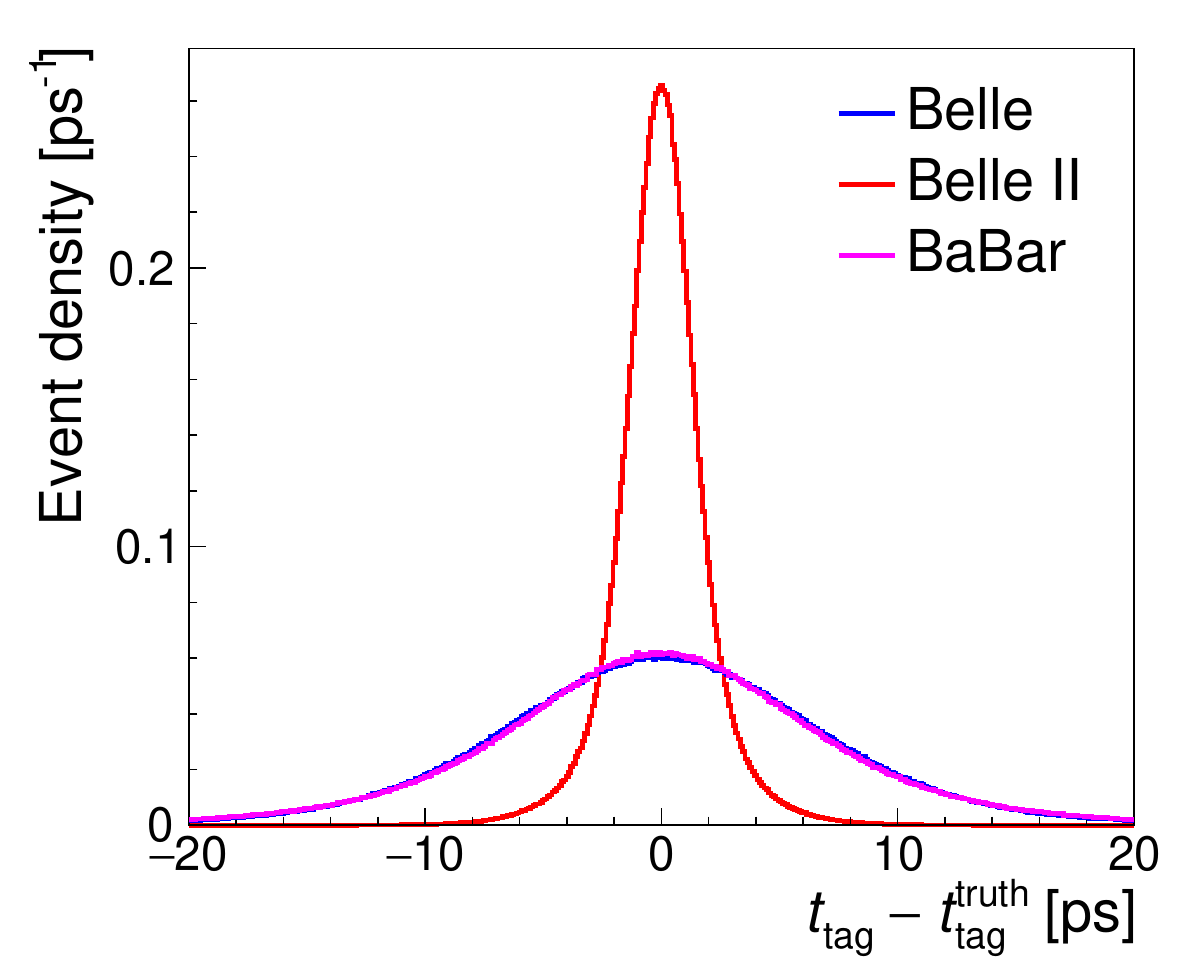}
\caption{Distribution of simulated $\ttag - \ttag^\mathrm{truth}$ obtained from the parameters of Table~\ref{tab:BfactoriesPars} for the (blue)~Belle-like, (red)~\belletwo-like, and (magenta)~BaBar-like  scenarios. The blue and magenta curves overlap.}
\label{fig:tTagRes}
\end{figure}

We then inspect the distributions of the differences $\ttag - \ttag^\mathrm{truth}$ between the \Btag decay times \ttag, reconstructed using Eq.~(\ref{eqn_chi2}), and the true decay times  $\ttag^\mathrm{truth}$ (Fig.~\ref{fig:tTagRes}). All distributions are approximately Gaussian with small exponential tails; their core widths provide estimates of the \ttag resolutions. The core width is approximately 1.3\,ps at \belletwo, significantly better than the 6.0\,ps figure achieved at previous-generation $B$~factories. To validate our simplified simulation, we also determine the \deltat resolutions from our model, and find Gaussian-core widths consistent with those observed at \belletwo, Belle, and BaBar. Nonetheless, our simplified simulation lacks non-Gaussian tails induced by the charmed-meson lifetimes for \Btag vertices from $B\to D$ decays. These tails can lead to an $\mathcal{O}(0.1)$ additional loss in sensitivity to mixing-induced \cp violation, which is irrelevant to the conclusions of our study. 

The better \belletwo\ performance compared with Belle and BaBar results from four main factors: (i) a smaller IR at SuperKEKB than at KEKB and PEP-II, and (ii) improved \Btag vertex resolution. Together, these lead to a smaller intersection between the IR ellipsoid and the \Btag tube, enhancing the precision in determining the \Btag production space-point. In addition, (iii) a nonzero beam-crossing angle, and (iv) a sufficient boost, jointly increase the fraction of \B mesons that exit the IR ellipsoid before decaying at \belletwo.\footnote{A smaller boost results in $B$ mesons being emitted with a wider angular distribution relative to the boost direction,  which reduces the uncertainty of the \Btag production point and improves $|\vec{L}|$ resolution. On the other hand, a smaller boost also deteriorates the time resolution by enlarging the denominator of Eq.~(\ref{eqn_sigmat_dep}).} 

%
\begin{figure*}[t]
\centering
\includegraphics[width=0.8\columnwidth]{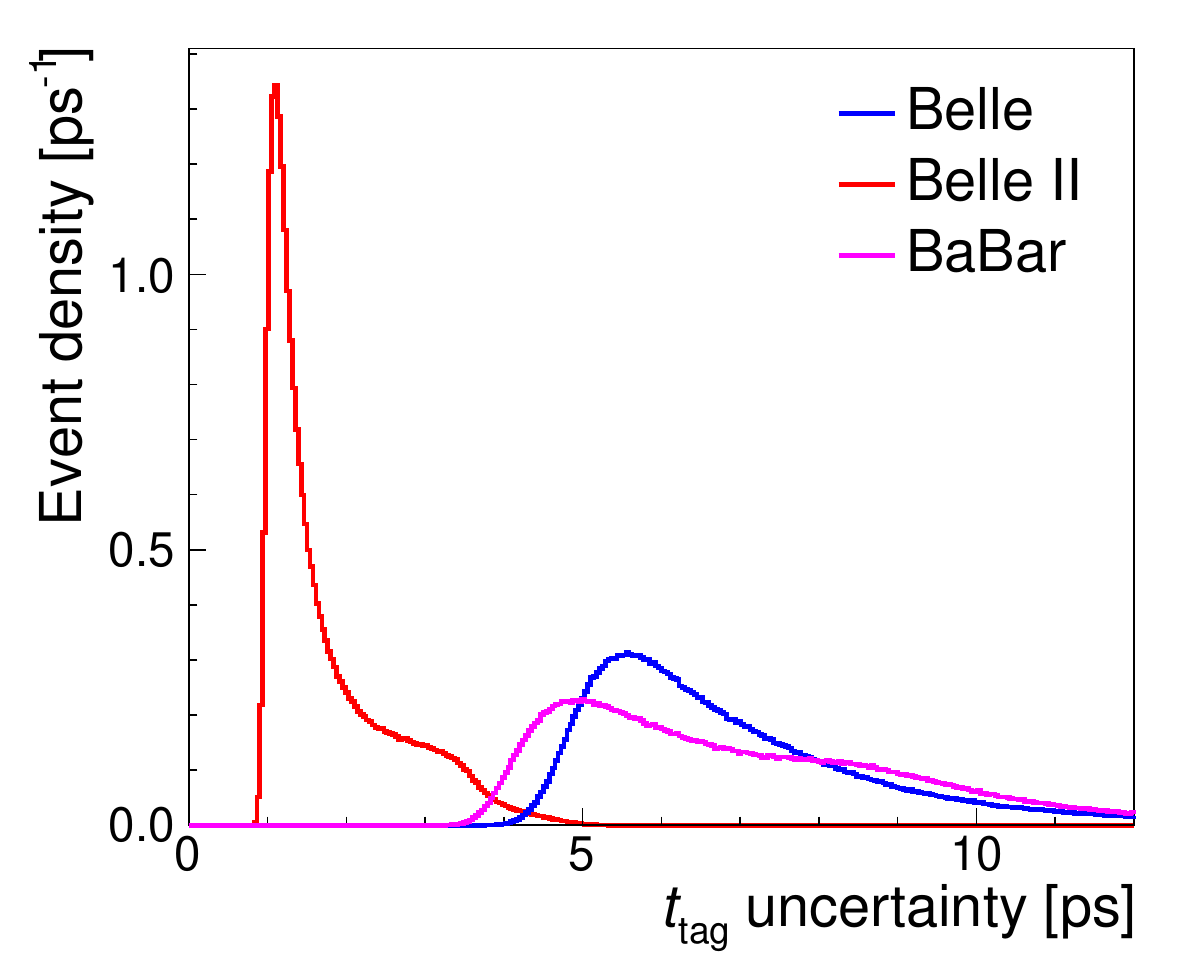}\hspace{1.5cm}
\includegraphics[width=0.83\columnwidth]{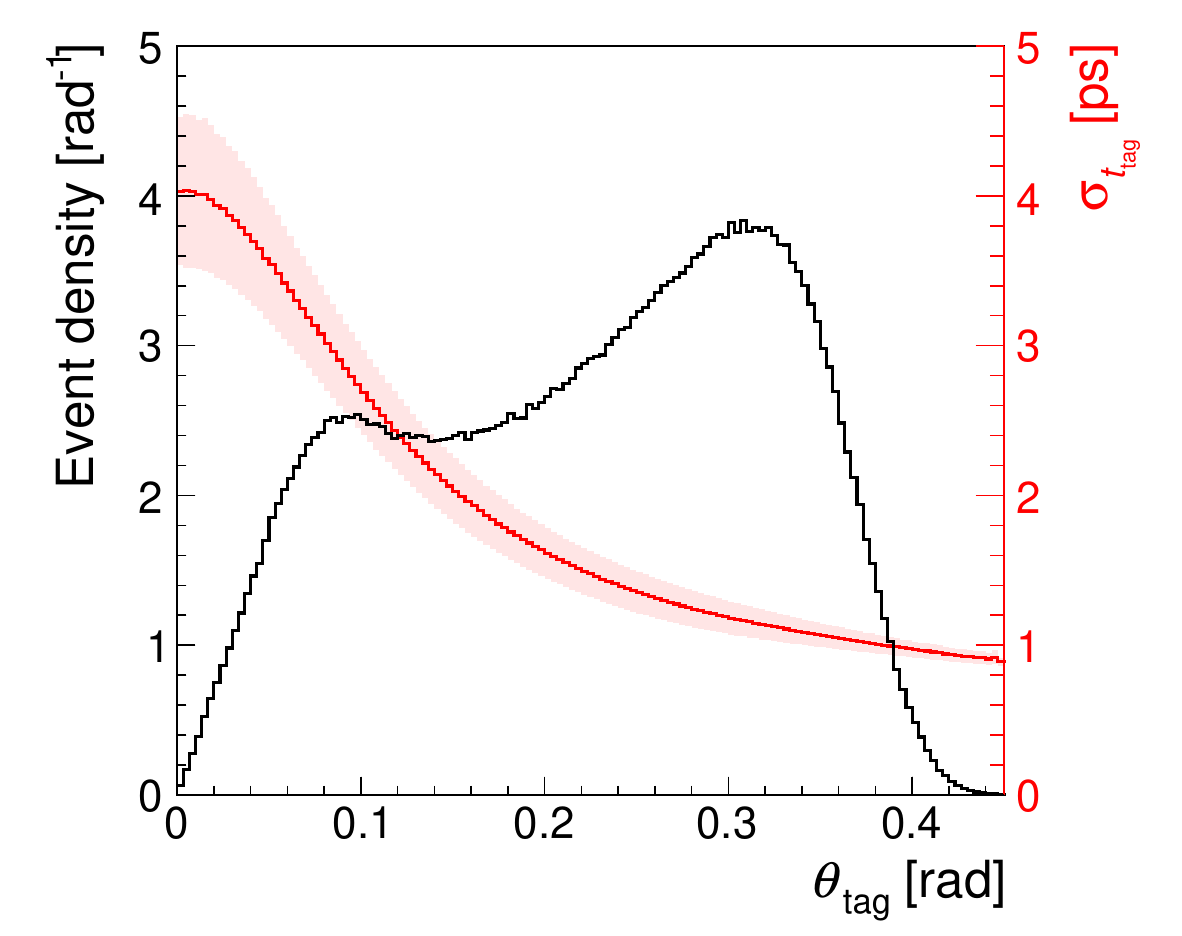}
   \caption{(Left panel) Expected distributions of $\sigma_{\ttag}$ for Belle~(blue), \belletwo~(red), and BaBar~(magenta), as obtained from the minimization of the $\chi^2$ function of Eq.~(\ref{eqn_chi2}). (Right panel) distributions of (black) the angle $\theta_{\rm tag}$ between the \Btag momentum and the $z$ axis and (red) $\sigma_{\ttag}$ as a function of that angle in a \belletwo-like scenario.  The shaded red band represents the 68.3\%  confidence interval.}
\label{fig:tTagSigmas}
\end{figure*}

The minimization of the $\chi^2$ in Eq.~(\ref{eqn_chi2}) also provides a determination of the expected \ttag uncertainty $\sigma_{\ttag}$, whose distributions are shown in Fig.~\ref{fig:tTagSigmas} (left). The resolution varies on an event-by-event basis yielding the distributions of  Fig.~\ref{fig:tTagRes} as integrals over all \ttag measurements.
When properly estimated, $\sigma_{\ttag}$ provides a means to distinguish between events with better or worse time resolution. We validate $\sigma_{\ttag}$ as a reliable estimate of the uncertainty on \ttag by examining the width of the distribution of $\ttag - \ttag^\mathrm{truth}$ as a function of $\sigma_{\ttag}$. We observe an accurate correspondence between $\sigma_{\ttag}$ and the width of the $\ttag - \ttag^\mathrm{truth}$ distribution, confirming the accuracy of the uncertainty estimate. The resolution $\sigma_{\ttag}$ can therefore be used  in fits to the time evolution, in analogy with per-event uncertainties in $\deltat$ for measurements of $\mathcal{A}_{\cp}(\deltat)$.

The uncertainty $\sigma_{\ttag}$ is independent from $|\vec{L}|$, due to the constraint from the \Btag momentum vector in Eq.~(\ref{eqn_xdec_fit}). It depends on the \Btag momentum-vector direction though (see Fig.~\ref{fig:tTagSigmas} (right)). 
The more $\vec{p}_\textrm{tag}$ is collinear with the main axis of the IR ellipsoid, the larger the intersection between the IR and the \Btag tube, hence the uncertainty on the production space-point and \ttag; the closer $\vec{p}_\textrm{tag}$ is to the transverse plane, the smaller the intersection and thus the uncertainty. The dependence on the direction generates the bimodal shape of the $\sigma_{\ttag}$ distribution in a \belletwo-like scenario. 
In the same scenario, minimizing the $\chi^2$ of Eq.~(\ref{eqn_chi2}) with the $\vec{p}_\textrm{tag}$ constraint improves the \ttag resolution by a factor of about three, compared to using the IR  center as the \Btag production space-point, {\it i.e.}, when \ttag is reconstructed directly by Eq.~(\ref{eqn_ttag_def}) without performing the vertex fit.

From the distribution of $\sigma_{t_\mathrm{tag}}$, we compute an effective time resolution, $\sigma^\textrm{eff}_{\ttag}$, which corresponds to the width of the single-Gaussian resolution that yields the same sensitivity on the \cp-violating coefficient $S$ as that obtained considering all values of $\sigma_{\ttag}$. We combine the uncertainties $\sigma_S(\sigma_{\ttag})$ on $S$, from the function in Fig.~\ref{fig:SCresScan}, weighted with  the normalized distribution $w(\sigma_{\ttag})$ of $\sigma_{\ttag}$, {\it i.e.}, 
\begin{equation}
    \frac{1}{\sigma_S^2(\sigma^\textrm{eff}_{\ttag})} = \int  \frac{w(\sigma_{\ttag})}{\sigma_S^2(\sigma_{\ttag})} d\sigma_{\ttag} \,.
\end{equation}
The resulting effective \ttag resolutions are 1.5\,ps at \belletwo, 6.3\,ps at Belle, and 6.0\,ps at BaBar. With respect to the case with perfect $\ttag$ resolution, the precision of $S$ is therefore expected to reduce by a factor of 1.7 for \belletwo, 8.0 for Belle, and 7.0 for BaBar, due to the time-resolution-dependent dilution of the asymmetry (see, {\it e.g.}, Fig.~\ref{fig:SCresScan}). 
These findings show that measurements of time-dependent \cp asymmetry without signal vertex only are feasible and competitive at \belletwo.

\section{Case study: $\Bz \to \piz\piz$ and impact on the CKM angle $\phi_2$}
To assess the potential impact quantitatively, we apply our method to a realistic use case, a measurement of time-dependent \cp-violating asymmetries in $\Bz\to\piz\piz$ decays at \belletwo.  If available, this measurement could significantly improve the determination of the CKM angle $\phi_2 = \arg(V_{td}V_{tb}^*/V_{ud}V_{ub}^*)$, where $V_{ij}$ are CKM matrix elements, enhancing in turn the constraining power of CKM-unitarity tests. Prior to using our approach to determine mixing-induced \cp violation in $B^0 \to \pi^0\pi^0$ decays, which is unknown, a  consistency check based on $B^0 \to J/\psi K^0$ decays would provide a high-precision validation.

The angle $\phi_2$ is typically constrained through a combination of results from decays related by isospin symmetry, $\Bz \to \pip\pim$, $\Bp \to \pip\piz$, and $\Bz \to \piz\piz$. This combination suppresses hadronic uncertainties due to penguin contributions in the $\Bz \to \pip\pim$ decay amplitude~\cite{Gronau:1990ka}. The current $4.5^\circ$ precision on $\phi_2$ is driven by the precision of an independent determination based on an analogous isospin analysis of $B \to \rho\rho$ decays. The $B \to \pi\pi$ analysis is less impactful also because it yields degenerate results due to an eight-fold ambiguity. The degeneracy originates from lack of experimental information on mixing-induced \cp violation in $\Bz\to\piz\piz$ decays, typically denoted with the $S_{00}$ coefficient. 
To date, a measurement of $S_{00}$ has  been considered feasible by only reconstructing the $\piz \to e^+e^-\gamma$ final state to enable signal-vertex reconstruction and the ensuing \deltat measurement. However, this standard approach requires datasets 50--100 times larger than those currently available~\cite{Ishino:2007pt,Belle-II:2018jsg}. We show that a measurement of $\mathcal{A}_{\cp}'(\ttag)$ yields highly constraining $S_{00}$ information by already using the data collected by \belletwo\ thus far.

We build upon the recent measurement of the $\Bz\to\piz\piz$ branching fraction and time-integrated \cp-violating asymmetry $C_{00}$, in which 126 signal decays were reconstructed in the \belletwo\ sample collected until 2022, and corresponding to an integrated luminosity of 362\,\invfb~\cite{Belle-II:2024baw}. 
We assume the same sample size and composition as in Ref.~\cite{Belle-II:2024baw}.  The sample is background-dominated: approximately 6850 events are from light-quark production processes (continuum) and 170 from other $B$ decays ($B\Bb$ background), such as $B^+ \to \rho^+(\to\pi^+\pi^0)\pi^0$, where the $\pi^+$ is not reconstructed. 

The results in Ref.~\cite{Belle-II:2024baw} were obtained from a fit to three background-discriminating observables  ($\Delta E$, $M_\mathrm{bc}$, and $C_t$) and one flavor-sensitive observable ($w_t$). The first observable is the difference between the $B$-candidate energy and half of the beam energy, $\Delta E= E_B^* - E_\mathrm{beam}^*$; the second observable is the $B$-candidate mass calculated from the $B$ momentum and the beam energy, $M_\mathrm{bc} = \sqrt{E_\mathrm{beam}^{*2} - p_B^{*2}}$; the third observable, $C_t$, is the transformed output of a classifier trained to suppress continuum; and the fourth observable, $w_t$, is the transformed fraction of incorrectly tagged events $w$, estimated by a flavor-tagging algorithm providing the \Btag flavor $q$~\cite{Belle-II:2024lwr}. All starred quantities are calculated in the center-of-mass frame. 

We generate simplified simulated data by sampling the probability density function (PDF) used in Ref.~\cite{Belle-II:2024baw}, extended to include the observables \ttag and $\sigma_{\ttag}$. Unlike in Ref.~\cite{Belle-II:2024baw}, we assume that the joint PDF factorizes as the product of the individual PDFs for each observable; in addition, we neglect flavor-tagging asymmetries and uncertainties on the flavor-tagging parameters. These approximations are not expected to bias appreciably the expected resolutions on the fit results, thus not compromising the validity of this study. The full PDF is 
\begin{equation}
\label{eq:pdf}
\begin{aligned}
&\mathcal{P}(\ttag,\sigma_{\ttag},\Delta E, M_\mathrm{bc}, C_t, w_t, q) = \\
& \sum_j f_j \mathcal{P}_j(\ttag, \sigma_{\ttag}; q, w) \mathcal{P}_j(\Delta E) \mathcal{P}_j(M_{\rm bc}) \mathcal{P}_j(C_t) \mathcal{P}_j(w_t),
\end{aligned}
\end{equation}
where $f_j$ is the fraction of component $j$, an index that indicate signal ($s$), continuum ($c$) or $B\Bb$ background ($b$), and $f_b$ equals $1 - f_s - f_c$.

The PDFs for $\Delta E$, $M_\mathrm{bc}$, $C_t$, and $w_t$ are taken from Ref.~\cite{Belle-II:2024baw}. Integrating Eq.~(\ref{eq:pdf}) over \ttag and $\sigma_{\ttag}$ yields the time-integrated model used in that reference. To validate our implementation, we fit to the simulated data generated from this model to determine signal fraction,  continuum fraction, and $C_{00}$. The fit yields unbiased results with Gaussian uncertainties. The average statistical uncertainty of $C_{00}$ equals $0.28 \pm 0.02$, consistent with the $0.30$ value reported in Ref.~\cite{Belle-II:2024baw}.

In the time-dependent analysis, the signal PDF derives from Eq.~(\ref{eqn_tTag}) after incorporating the effect of the wrong-tag fraction $w$, and the convolution with a Gaussian time-resolution model $\mathcal{R}_{\sigma_{\ttag}}$, 
\begin{linenomath}
\begin{align}\label{eqn_tTagFull}
\begin{split}
&\mathcal{P}_s(\ttag, \sigma_{\ttag}; q, w) =  \\
&\frac{e^{-\ttag'/\tau}}{2\tau} \biggl(1 + 
q (1-2w) 
\bigl[ \scp' \sin \Delta m^{} (\ttag'-\hat{t})\\
&-\ccp'\cos \Delta m^{} (\ttag'-\hat{t}) \bigr] \biggr) \otimes \mathcal{R}_{\sigma_{\ttag}}(\ttag - \ttag')\,.
\end{split}
\end{align}
\end{linenomath}
Here, $\sigma_{\ttag}$ indicates the per-event tag decay-time resolution, which effectively increases the importance of events with better resolution in the fit. The physical \cp-violating parameters $S_{00}$ and $C_{00}$ are related to $S'$ and $C'$ via Eq.~(\ref{eqn_SC}). 

For backgrounds, we use a $\delta$-function distribution for continuum, and an exponential decay with the $B^+$ lifetime for the $B\Bb$ component. This is motivated by the dominance of the $B^+ \to \rho^+ \pi^0$ contribution reported in Ref.~\cite{Belle-II:2024baw}. Both are convolved with the same Gaussian resolution as the signal. We assume the $\sigma_{\ttag}$ distribution is the same across components.

We simulate three scenarios, (i) the 365\,\invfb dataset that \belletwo\ collected up to 2022, (ii) an intermediate sample corresponding to 5\,\invab, and (iii) the design full dataset of 50\,\invab. These assess the feasibility, mid-term reach, and ultimate sensitivity, respectively,  of our approach in a \belletwo-like experiment. In each case, simulated data are generated assuming $S_{00} = 0.65$, which is the value favored by an indirect $\phi_2$ determination~\cite{Charles:2017evz}, and $C_{00} = 0.00$, which is compatible with current determinations from time-integrated measurements and allows assessing more directly the impact on mixing-induced \cp violation.   
For each scenario, we fit simulated data to determine the signal and continuum fractions, $S_{00}$, and $C_{00}$. 

\begin{figure}
\includegraphics[width=0.9\columnwidth] {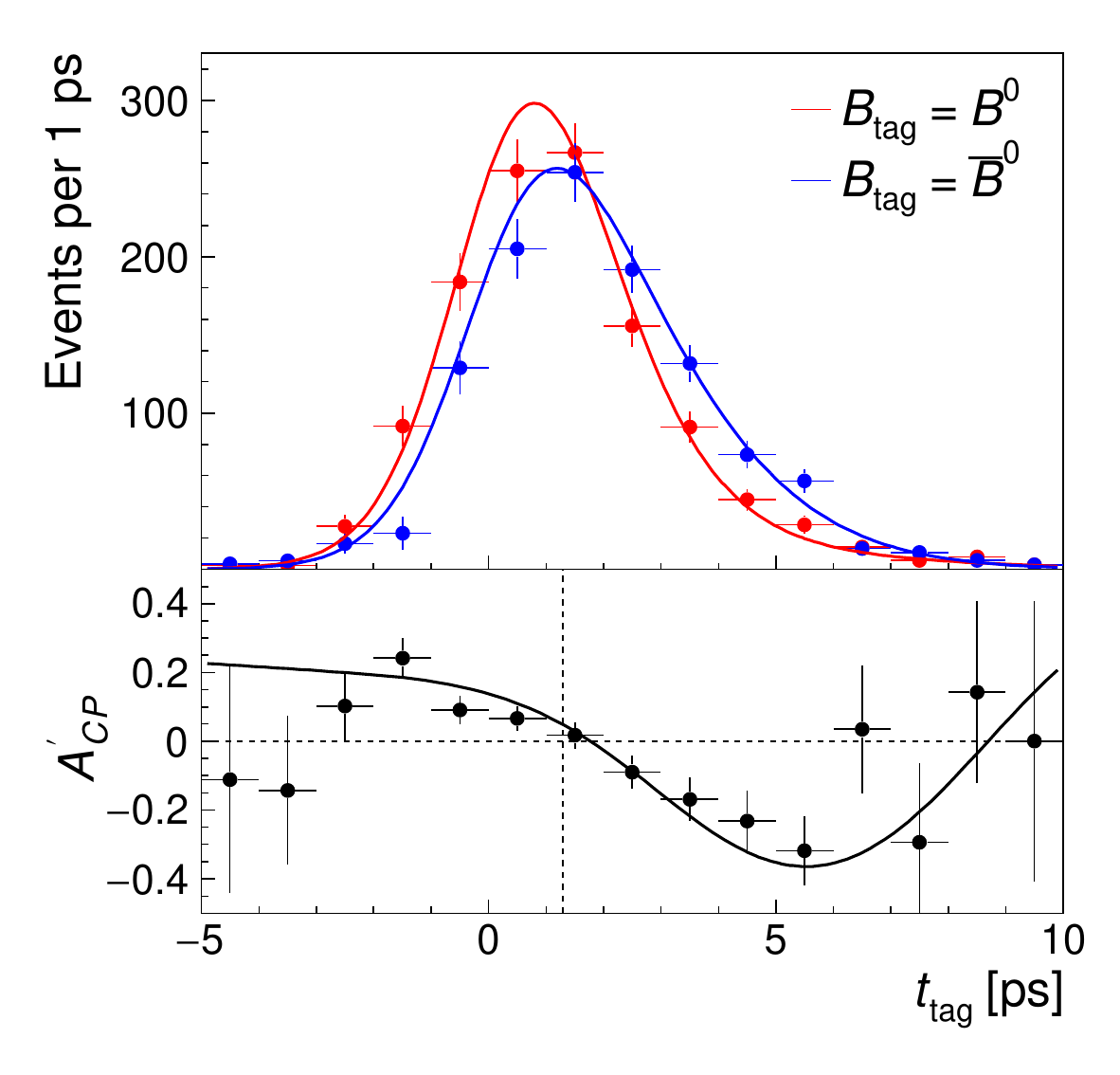} 
\caption{(Top panel) example of a background-subtracted \ttag distributions for (red) $B^0$ and (blue) $\overline{B}^0$ mesons and (bottom panel) corresponding decay-rate asymmetry in a \belletwo-like 50\,\invab scenario, with fit projections overlaid. Simulated data are displayed in the signal-enhanced region $-0.10 < \Delta E < 0.05$\,GeV, $5.275 < M_\mathrm{bc} < 5.285$\,GeV, $C_t > 0$.
In addition, we apply $w < 0.2$ and $\sigma_{\ttag} < 2$\,ps for enhancing visualization of the asymmetry. Corresponding fit results are $S_{00} = 0.69\pm 0.06$ and $C_{00} = 0.00\pm 0.02$.}
\label{fig:fit_projection_example}
\end{figure}

\begin{table}[b]
\begin{tabular}{cccccc}
\hline\hline 
Sample size   & \phantom{0}& \multicolumn{2}{c}{\ttag} & \phantom{0} & \deltat \\
    $[\invab]$      & \phantom{0}& $\sigma_{S_{00}}$  & $\sigma_{C_{00}}$ & \phantom{0}& $\sigma_{S_{00}}$  \\ 
\hline
$0.36$        & \phantom{0}&  $0.75$         &        $0.27$  & \phantom{0} & --  \\
$5$        & \phantom{0}&  $0.18$         &        $0.07$  & \phantom{0} & --  \\
$50$        & \phantom{0} &  $0.06$        &        $0.02$ & \phantom{0} & $0.28$  \\
\hline\hline
\end{tabular}
\caption{Expected statistical uncertainties on $S_{00}$ and $C_{00}$ from \ttag-dependent analyses in the various \belletwo-like scenarios, compared with the $S_{00}$ uncertainty expected in the standard \deltat analysis~\cite{Belle-II:2018jsg}. 
}
\label{tab:projectedUnc}
\end{table}

An example of the background-subtracted \ttag distribution for \Bz- and \Bzb-tagged decays and the corresponding asymmetry, with fit projections overlaid, is shown in Fig.~\ref{fig:fit_projection_example}. All estimates are unbiased and have Gaussian uncertainties. The resulting average statistical uncertainties for $S_{00}$ and $C_{00}$ are listed in Table~\ref{tab:projectedUnc}. Simulation shows that the sensitivities to $S_{00}$ and $C_{00}$ do not depend on the assumed values of these parameters. In that table, we also compare our results with those from the standard method using \deltat from $e^+e^-\gamma$ vertexing, following Ref.~\cite{Belle-II:2018jsg}.  The projected uncertainty on $S_{00}$ from our method for the ultimate sample size scenario is 0.06, approximately five times smaller than the 0.28 reported in Ref.~\cite{Belle-II:2018jsg} for the standard method. Our approach achieves the same sensitivity with approximately 2\,\invab, which would be a 20-times smaller dataset.

These findings show promising potential for the determination of $\phi_2$. We assess the potential impact from a $B \to \pi\pi$ isospin analysis that incorporates our projected results in addition to the existing inputs
for three cases: (i) using all $B \to \pi\pi$ measurements available to date; (ii)  combining all $B \to \pi\pi$ measurements available to date with the $S_{00}$ results expected from our method in a \belletwo-like scenario based on the 0.36 ab$^{-1}$ sample collected up to 2022; and
(iii) combining all $B \to \pi\pi$ measurements available to date with
the projected $S_{00}$ determination for a \belletwo-like scenario with 5\,\invab.
We use values~of branching fractions for $B^0 \to \pi^+\pi^-$, $B^+ \to \pi^+\pi^0$, and $B^0 \to \pi^0\pi^0$ decays, $B$-meson lifetimes, and the \cp-violating coefficients $S_{+-}$, $C_{+-}$, $S_{00}$, and $C_{00}$ as inputs, following the analysis of Ref.~\cite{Gronau:1990ka}. We use $S_{00} = 0.65 \pm 0.75$ ($0.65 \pm 0.18$) in the  0.36 (5)\,ab$^{-1}$ sample for the unknown value of the mixing-induced \cp violation coefficient, in which the central value follows Ref.~\cite{Charles:2017evz} and the uncertainties are from Table~\ref{tab:projectedUnc}.
We use known values for all other inputs~\cite{Banerjee:2024znd}. The \belletwo\ results in Ref.~\cite{Belle-II:2024baw} are assumed uncorrelated with $S_{00}$.
While the details of the $\phi_2$ results depend moderately on these choices, the general conclusions are expected to hold.   To assess the sole impact of our method, all inputs except for $S_{00}$ are kept unchanged in all scenarios, although the precision on most of them is expected to improve as well.

\begin{figure}[t]
\vspace{3mm}
    \makebox[0.49\textwidth]{\includegraphics[width=0.49\textwidth]{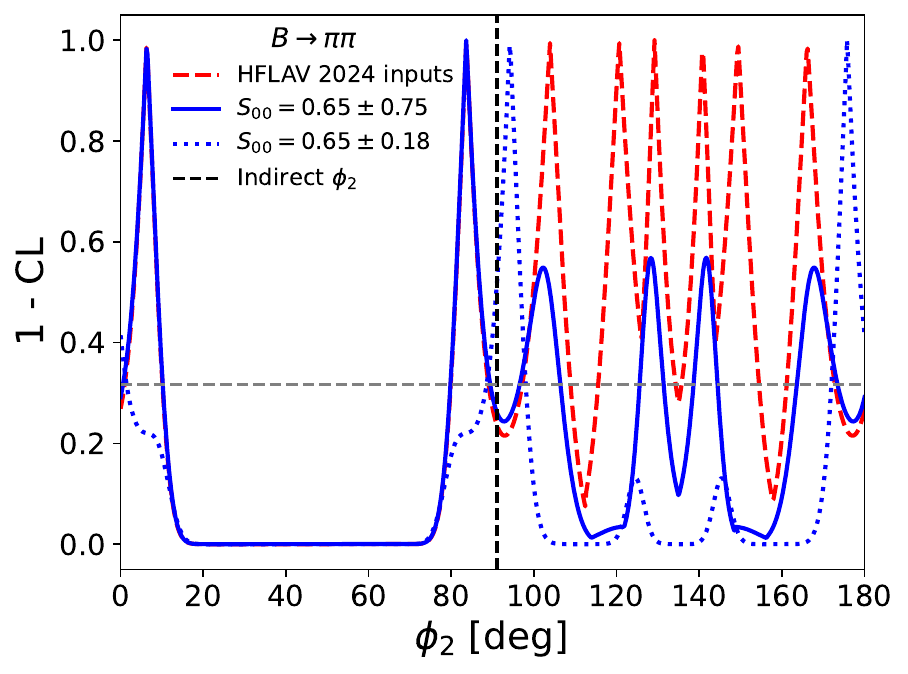}}
    \caption{$P$-value as a function of the CKM angle $\phi_2$ from an isospin-based combination of (red dashed) all currently available $B \to \pi\pi$ results only, (blue solid) all currently available results and $B^0 \to \pi^0 \pi^0$ results expected from this approach in a \belletwo-like scenario based on the 0.36 ab$^{-1}$ sample collected up to 2022, and (blue dashed) all current results and \belletwo-like  $B^0 \to \pi^0 \pi^0$ results expected from this approach in an hypothetical 5\,ab$^{-1}$ sample. 
    All other inputs are the latest known values~\cite{Banerjee:2024znd}.}
    \label{fig:alpha_pipi}
\end{figure}

Figure~\ref{fig:alpha_pipi} shows the results of the isospin analysis. Resulting  $p$-values as functions of $\phi_2$ are displayed.  
Already with the \belletwo\ sample collected up to 2022, the results enabled by our method would reduce the solution degeneracy from eight solutions to two, and would reduce the 68\% confidence-level interval around the solution favored by global fits. 
This demonstrates that a \ttag-based time-dependent analysis of $B^0 \to \pi^0\pi^0$ significantly improves the global determination of $\phi_2$. 

\section{Summary}
We develop a method that enables measurement of time-dependent \cp-violation in $\Bz$ mesons produced in energy-asymmetric electron-positron collisions near the \FourS resonance and reconstructed without a signal vertex. The relevant time-evolution is expressed solely as a function of the decay time of the pair-produced tag-$B$ meson and information on signal \cp violation is provided by the quantum correlation of the  $\Bz\Bzb$ pair.  Key is a compact \epem interaction region and excellent vertex resolution, which enable precise measurement of the tag-$B$ decay time.
This approach extends and enhances sensitivity to time-dependent \cp-violating parameters in decays without charged particles, such as those involving only $\piz$, $\Kz$, and photons in the final state. When applied to the $\Bz \to \piz\piz$ decay, for instance, simulation shows that the method achieves a sensitivity on mixing-induced \cp violation that would require 20-times larger samples with standard approaches. This implies that meaningful constraints on the \cp-violating coefficients $S$ and $C$ can already be achieved for $\Bz \to \piz\piz$ decays with the current 0.5 ab$^{-1}$ Belle~II sample. The expected sensitivity on $S$ would reduce the degeneracy of solutions for the CKM angle $\phi_2$ in the isospin analysis of $B \to \pi\pi$ decays, substantially improving its determination. 
These findings establish the $\ttag$-dependent analysis as a powerful enabler for precision \cp-violation studies at \belletwo\ and beyond.

\appendix

\section{Formulas for tag-side vertexing}
\label{app_formulas}
The linearity of the $\chi^2$ regression of Eq.~(\ref{eqn_chi2}) allows to derive explicit analytic expressions for the best-fit vertex positions, \ttag times, and  $\sigma_{t_\mathrm{tag}}$ uncertainties.
We introduce the vector $\vec{n} = \vec{p}_\mathrm{tag} / m_B$ to simplify the algebra along with
\begin{align}
\begin{split}
E &= ( V_\mathrm{IR}^{-1} + V_\textrm{tag}^{-1} )^{-1}\,, \\
\vec{x}_m &= E (V_\mathrm{IR}^{-1} \vec{x}_\mathrm{IR} + V_\textrm{tag}^{-1} \vec{x}_\textrm{tag})\,.
\end{split}
\end{align}
We obtain the tag decay time
\begin{align}
\begin{split}
t_\mathrm{tag} = \frac{ \vec{n}^{\,T} V_\textrm{tag}^{-1} (\vec{x}_m - \vec{x}_\textrm{tag})} {\vec{n}^{\,T} ( V_\textrm{tag}^{-1} E V_\textrm{tag}^{-1} - V_\textrm{tag}^{-1}) \vec{n} }\,, \label{eq_tTagFormula}
\end{split}
\end{align}
and the production and decay vertices determined from the fit are then expressed as
\begin{align}
\begin{split}
\vec{x}_\mathrm{prod} &= \vec{x}_m - t_\mathrm{tag} E V_\textrm{tag}^{-1} \vec{n}\,, \\
\vec{x}_\mathrm{dec} &= \vec{x}_\mathrm{prod}  + t_\mathrm{tag}  \vec{n}\,.
\end{split}
\end{align}
The tag decay time calculated using $t_\mathrm{tag} = m_B\, \vec{p}_\mathrm{tag} \cdot (\vec{x}_\mathrm{dec} - \vec{x}_\mathrm{prod}) /  |\vec{p}_\mathrm{tag}|^2$ is identical to the value resulting from Eq.~(\ref{eq_tTagFormula}).  

The inverse variance of \ttag as determined from the fit is 
\begin{align}
\begin{split}
\frac{1}{\sigma_{t_{\textrm{tag}}}^2} &= (E V_\textrm{tag}^{-1} \vec{n})^T V_\mathrm{IR}^{-1} (E V_\textrm{tag}^{-1} \vec{n})\\
&+ (E V_\textrm{tag}^{-1} \vec{n} - \vec{n})^T V_\textrm{tag}^{-1} (E V_\textrm{tag}^{-1} \vec{n} - \vec{n})
\end{split}
\end{align}
In the limit of isotropic $B_\mathrm{tag}$ vertex resolution $\sigma_\textrm{tag}$ and an elongated interaction region, the $t_\mathrm{tag}$ resolution approximates to
\begin{equation}
\sigma_{t_\mathrm{tag}} \approx \frac{ \sqrt{\sigma_{{\rm IR}_{xy}}^2 + \sigma_\textrm{tag}^2}} { (\beta\gamma)_{\mathrm{tag}}\, \sin \theta_\mathrm{tag}},
\label{eqn_sigmat_dep}
\end{equation}
where $\sigma_{{\rm IR}_{xy}}$ is the transverse size of the interaction region,  $(\beta\gamma)_{\mathrm{tag}}$ is the boost of the $B_\mathrm{tag}$ meson, and $\theta_\mathrm{tag}$ is the angle between $\vec{p}_\mathrm{tag}$ and the main axis of the interaction region.

\section*{Acknowledgments} 
We thank A. Di Canto, T. Gershon, and J. Libby for useful suggestions on the manuscript.

\bibliographystyle{apsrev4-2-mod}
\bibliography{references}

\end{document}